\documentclass{iau-JDSS}
\usepackage{graphicx}

\title[NEI Modelling of the ISM \& Hausdorff Dimension] 
{NEI Modelling of the ISM - Turbulent Dissipation \& Hausdorff Dimension}

\author[Avillez \& Breitschwerdt]   
{Miguel A. de Avillez$^1$ \and Dieter Breitschwerdt$^2$}

\affiliation{$^1$Department of Mathematics, University of \'Evora, 7000 \'Evora, Portugal \break email: mavillez@galaxy.lca.uevora.pt\\[\affilskip]
$^2$Department of Astronomy \& Astrophysics, Technical University of Berlin, D-10623 Berlin, Germany \break email: breitschwerdt@astro.physik.tu-berlin.de}

\pubyear{2010}
\volume{Volume 15}  
\pagerange{119--126}
\date{?? and in revised form ??}
\jname{Highlights of Astronomy, Volume 14}
\editors{Ian F Corbett, ed.}
\begin{document}

\maketitle

\begin{abstract}
High-resolution non-ideal magnetohydrodynamical simulations of the turbulent magnetized ISM, powered by supernovae types Ia and II at Galactic rate, including self-gravity and non-equilibriuim ionization (NEI), taking into account the time evolution of the ionization structure of H, He, C, N, O, Ne, Mg, Si, S and Fe, were carried out. These runs cover a wide range (from kpc to sub-parsec) of scales, providing resolution independent information on the injection scale, extended self-similarity and the fractal dmension of the most dissipative structures.
\end{abstract}

\firstsection 
\section{Introduction}
In star forming disk galaxies, matter circulation between stars and the interstellar medium (ISM), in particular the energy input by supernovae, determines the dynamical and chemical evolution of the ISM, and hence of the galaxy as a whole. So far ISM models used radiative cooling assuming collisional ionization equilibrium (CIE), which is a good approximation providing the cooling time is much longer than the recombination times (e.g., Kafatos 1973). A condition verified for most of the ions for temperatures $> 10^{5.8}$ K. At lower temperatures departures from equilibrium are expected. While in CIE the ionization fractions depend only on the temperature and are sharply peaked, in NEI these same fractions depend on the dynamical and thermal history of the plasma. These departures affect the local cooling, which is a time-dependent process that controls the flow dynamics, feeding back to the thermal evolution by a change in the density and internal energy distribution, which in turn modifies the thermodynamic path of non-equilibrium cooling (Breitschwerdt \& Schmutzler 1999).

\section{ISM Modelling}
Following Avillez \& Breitschwerdt (2007) we study the evolution of the ISM in a patch of the Galaxy with $0\leq (x,y)\leq 1$ kpc size in the Galactic plane, and $\left|z\right|\leq 10$ kpc perpendicular to it using non-ideal MHD equations coupled to the time-dependent calculation of the ionization structure of the plasma. The adopted model includes: (i) SNe types Ia and II occurring at the Galactic SN rate - 40\% of the SNe II occur in the field (ii) gravitational field of the stellar disk, (iii) local self-gravity, (iv) heat conduction, (v) backgroung heating due to the UV photon field, (vi) mean (of $3.0 \mu$ G) and turbulent magnetic field components corresponding to a total field of 4.5 $\mu$ G, and (vii) radiative cooling calculated on the spot and using solar abundances of Asplund et al. (2005). The NEI calculation takes into account the time-dependent ionization structure of the 10 most important elements in nature - H, He, C, N, O, Ne, Mg, Si, S, and Fe - and includes collisional ionization by thermal electrons, autoionization, charge exchange reactions, radiative and dielectronic recombination, photoionization, comptonisation, Auger effect and ionization of H and He by photo- and Auger electrons. 
\section{Fractal Dimensions}

\begin{figure}[thbp]
\centering
\includegraphics[width=0.4\hsize,angle=-90]{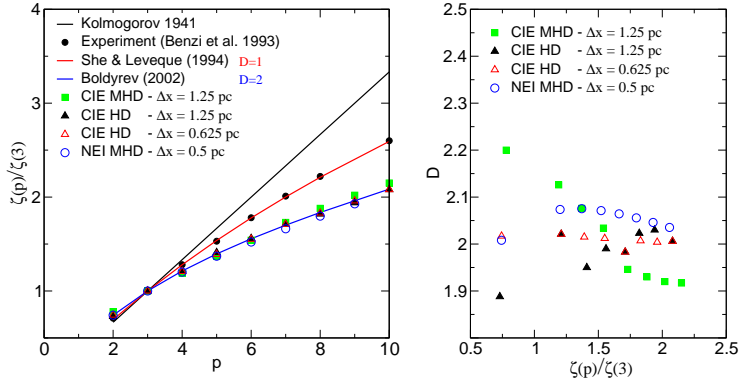}
\caption{\emph{Left:} The scalings $\zeta(p)/\zeta(3)$ vs. order $p$. Black, red and blue lines refer to Kolmogorov (1941), She-Lev\'eque (1994) and Boldyrev (2002) models, respectively; Bullets refer to data of Benzi et al. (1993); Triangles (black and red) and green squares refer to CIE HD (1.25 pc and 0.625 pc resolutions) and MHD (1.25 pc resolution) runs of Avillez \& Breitschwerdt (2007); Blue circles correspond to NEI MHD (0.5 pc resolution) run discussed here. \emph{Right:} The Hausdorff dimension $D$ of the most dissipative structures vs. $\zeta(p)/\zeta(3)$, with $p=2,4,...,10$ shown in the left panel.
\label{scalings} }
\end{figure}
We measured the velocity structure functions $\langle \delta v_{l}^{p}\rangle$ and determined the corresponding scalings $\zeta(p)/\zeta(3)$ and the Hausdorff dimension $D$ of the most dissipative structures (Figure~\ref{scalings}). $D$ oscillates around 2, indicating that turbulent energy is dissipated preferentially through shocks in the HD cases and a combination of shocks and current sheets in the MHD runs. In the presence of the magnetic field (ideal and non-ideal calculations) $D$ departures from 2 due to the anisotropy induced by the magnetic field. The NEI and CIE runs have similar fractal dimensions for the most dissipative structures and reproduce experimental and numerical results previously published.

\begin{acknowledgments}
M.A would like to thank the organizers for the invitation.
\end{acknowledgments}

\end{document}